\documentclass[12pt]{article}
\begin{document}

\title{\bf A Theological Argument for an Everett Multiverse}

\author{
Don N. Page
\thanks{Internet address:
profdonpage@gmail.com}
\\
Department of Physics\\
4-181 CCIS\\
University of Alberta\\
Edmonton, Alberta T6G 2E1\\
Canada
}

\date{2012 December 21\\
Mayan Long Count Calendar 13.0.0.0.0,\\
or day 1,872,000, end of the 13th b'ak'tun}

\maketitle
\large
\begin{abstract}
\baselineskip 21 pt

Science looks for the simplest hypotheses to explain observations.  Starting with the simple assumption that {\em the actual world is the best possible world}, I sketch an {\it Optimal Argument for the Existence of God}, that the sufferings in our universe would not be consistent with its being alone the best possible world, but the total world could be the best possible if it includes an omnipotent, omniscient, omnibenevolent God who experiences great value in creating and knowing a universe with great mathematical elegance, even though such a universe has suffering.

God seems loathe to violate elegant laws of physics that He has chosen to use in
His creation, such as Maxwell's equations for electromagnetism or Einstein's
equations of general relativity for gravity within their classical domains of
applicability, even if their violation could greatly reduce human suffering
(e.g., from falls).  If indeed God is similarly loathe to violate quantum
unitarity (though such violations by judicious collapses of the wavefunction
could greatly reduce human suffering by always choosing only favorable
outcomes), the resulting unitary evolution would lead to an Everett multiverse
of `many worlds', meaning many different quasiclassical histories beyond the
quasiclassical history that each of us can observe over his or her lifetime. 
This is a theological argument for one reason why God might prefer to create a
multiverse much broader than what one normally thinks of for a history of the
universe.

\end{abstract}

\normalsize

\baselineskip 25 pt

\newpage

Humans and especially scientists tend to favor the simplest explanations for observations, as expressed by such principles as Occam's Razor.  For example, currently physicists are searching for a mathematically elegant complete theory of the universe, with one candidate partial theory being superstring/M theory.

One may attempt to quantify the plausibility of competing candidate theories by a Bayesian analysis in which one initially weights theories by their prior probabilities (logically before considering the observations) but then multiplies these prior probabilities by what is somewhat misleadingly called the likelihoods of the theories, which are the conditional probabilities of one's particular observation given each of the theories (that is, assuming in turn that each theory is correct).  When this product of the prior probability and the likelihood for each theory is normalized by dividing by the sum of the products for all theories, one gets the posterior probability for the theory, the probability that it is correct given the observation.

Of course, this is a highly idealized procedure, since we do not have even one
single complete plausible theory enabling us to calculate the probability of
one's observation.  Also, the prior probabilities appear to be unavoidably
subjective.  Even if we make the subjective choice to use Occam's razor and
assign higher prior probabilities to simpler theories, it is subjective which
theories are simpler.  Furthermore, even if one did produce an ordering of
theories, say with the $n$th theory being simpler than the $(n+1)$th for each
positive integer $n$, there would still be the subjectivity of how to choose a
set of prior probabilities that decrease with $n$.  (A simple choice one could
make for a countably infinite set of theories ordered in increasing complexity
would be prior probabilities that are $1/2^n$, but there are infinitely many
other algorithms for constructing an infinite sequence of decreasing real
nonnegative numbers that sum to one as probabilities should.)

Although it is subjective how to assign the prior probabilities, and we do not yet know how to calculate the likelihoods for any plausible complete theory (one that would predict the probabilities of all possible observations), there does seem to be a trade-off between the priors and the likelihoods.  For example, the simplest theory I can think of (to which I might assign a prior probability of $1/2$) is the theory that nothing concrete exists.  However, this gives zero conditional probability for one's observation, so its posterior probability is zero.  The second-simplest theory I can think of is that all possible observations exist with equal probability (to which I might assign a prior probability of $1/2^2 = 1/4$), but since presumably there are an infinite number of logically possible observations, the likelihood of that theory, the probability of one's particular observation. would be $1/\infty = 0$, so this theory would also have a posterior probability of zero.

At the other extreme, the maximum-likelihood theory would be the one that gives unit probability to one's observation.  Since the probabilities for observations given by a theory must sum to one, this theory must give zero probability for all other observations, so it is an extreme solipsistic theory, saying not only that no one else's observations exist, but also that no other observations exist for the one in question at any other time.  We cannot logically rule out such a maximum-likelihood theory, but I suspect that one's observation is sufficiently complex that any complete theory predicting it and only it with positive probability would have to be sufficiently complex that it would be reasonable (though still subjective) to assign that theory such an extremely low prior probability that other complete theories exist (though presently unknown in complete detail) that would reasonably be assigned sufficiently higher prior probabilities that the products of those priors with the (necessarily lower, and probably quite a lot lower) likelihoods of those theories would still be higher than the product of the prior and the unit likelihood of the extreme solipsistic maximum-likelihood theory.

So neither maximal simplicity nor maximum likelihood seems plausible in the sense of giving a theory with nearly the maximum posterior probability.  In physics, one might imagine that a theory with high posterior probability is a fairly elegant mathematical theory, such as superstring/M theory is hoped to be (though no one knows a complete version of even just the dynamical part of this theory, which would have to be augmented with a specification of the quantum state and by rules for extracting the probabilities of observations from the quantum state in order to qualify as a complete theory; a so-called `Theory of Everything' or TOE that just gives the dynamical laws is certainly not a complete theory for the universe).  

However, even if such a theory were correct, it might be considered to leave it still unexplained why this rather simple, but not maximally simple, theory is correct rather than an even simpler theory.  One possibility is that there is an even simpler principle of why the world is as it is.  Although we humans have a very poor grasp on it (and so usually find a mathematical theory easier to calculate with), one principle that could conceivably be a simpler explanation is the following hypothesis about the world (the entirety of all that exists):

\hspace*{1.0in} {\em The actual world is the best possible world.}

Let me explore my dim understanding of this hypothesis and consider what it might plausibly be taken to imply, eventually leading to the question of whether or not there is an Everett multiverse.  I am not claiming that the steps in this argument are by any means logically complete, but I am offering them as a very crude sketch for what seems to me as if it might be a plausible picture of the world.

By hypothesizing that the actual world is the best possible world, I mean that in some sense it maximizes value.  I take the value that is maximized to be the intrinsic value of conscious or sentient experiences.  A painting may have instrumental value in eliciting positive feelings in a viewer, but I take the intrinsic value to be the pleasure or happiness or joy (or whatever positive word one might use) of the sentient experience of the viewer.  Assuming that the painting itself has no conscious awareness, I do not ascribe to it any intrinsic value.  Similarly, the laws of physics are presumably not sentient, but I do believe that they have great instrumental value in leading to pleasure in conscious experiences.
	
So what I mean by intrinsic value excludes entities without consciousness (which can only have instrumental value).  However, I do not assume that entities with consciousness have only intrinsic value, the value of their own conscious experiences.  Humans can be highly instrumental in leading to intrinsic value in the happiness of the conscious experiences of others.  Indeed, usually when we speak of a good person, we are not speaking of the intrinsic value of his or her own sentient experiences, but of the way his or her actions lead others to have positive conscious experiences.  We do not say Hitler was a good man even if he experienced happy thoughts while dancing a jig during World War II.  The suffering he caused is sufficient reason to label him as evil, despite what his own experiences may have been.
	
But when I consider the total goodness of the world, I do mean intrinsic value---pleasure minus suffering, happiness minus unhappiness, joy minus agony---of all conscious experiences that occur.  I am supposing that in our actual world, this total goodness is the greatest that is logically possible.

Now I have personally experienced a lot of happiness in my life, and I have seen many other people appear to be happy, so I do think that our universe does have considerable positive intrinsic value.  But there is also much suffering and unhappiness, so it does not seem very plausible that the intrinsic happiness within this universe alone is the maximum logically possible.

However, besides the mixed happiness that our universe contains, our universe seems to have a very high degree of mathematical elegance and beauty.  Humans can partially appreciate some of this elegance, so it does seem to increase the intrinsic value of the universe.  On the other hand, most of us (even we diehard physicists) would probably be happier if disasters, diseases, and cruelty were eliminated from the universe, even at the cost of less mathematical elegance and beauty for the laws of physics.

But if one considers the possibility that there may be a Being outside the universe that has an enormous appreciation for the mathematical elegance of the universe, and perhaps an even greater appreciation if that Being were the Creator of the universe, then it seems that the hypothesis that the actual world is the best possible world might plausibly lead to what I call the {\it Optimal Argument for the Existence of God} as such a Being outside the universe who created and experiences tremendous value in having much greater knowledge and appreciation for the mathematical elegance of the universe than we creatures within the universe can have.  

One might by further considerations of simplicity postulate that such a Being not only has very great knowledge of the universe but actually has all possible knowledge, so that such a Being is omniscient.  Considerations of simplicity might also suggest that such a Being is not only capable of creating our universe but also can do anything logically possible that His nature desires to do, so that such a Being is omnipotent.  And under our original assumption that the actual world is the best possible world, it seems plausible to assume that such a Being is all-good or omnibenevolent.  Of course, I am not claiming that I have any logical proof of the existence of such a Being or of these properties of omniscience, omnipotence, and omnibenevolence, but they do seem to me not unreasonable assumptions to make if one accepts the simple hypothesis that the actual world is the best possible world.  The hypothesis that the actual world is the best possible world certainly appears to me to be more consistent with a world with an omniscient, omnipotent, omnibenevolent Being (which I shall henceforth call God, since these are the main properties traditionally assigned to God), than with a world in which our universe stands alone, with its mathematical elegance not fully appreciated.

In particular, if there were no God who is a Person (at least in the sense of having conscious experiences), and if our universe (perhaps a multiverse) were the only entity with any conscious experiences within it, then it would seem that such a world could be better if all the conscious experiences in our universe were happy.  This might not be consistent with the actual laws of our universe, but surely it is logically possible and could be the case if the laws were suitably different from what they actually are.  Therefore, under the assumption of no personal God, our observations would seem to be incompatible with the hypothesis that the world is the best possible.  (Of course, atheists rarely do make the hypothesis that the world is the best possible, though even Richard Dawkins has said, ``The world and the universe is an extremely beautiful place, and the more we understand about it the more beautiful does it appear.''\footnote{http://en.wikiquote.org/wiki/Richard\_Dawkins.})  Despite my appreciation as a physicist for the beautiful laws of physics, I myself would much prefer to give up these laws whenever necessary to prevent cancer in people I know, or earthquakes in places like our beloved Haiti.
	
On the other hand, if there is an all-knowing God who is completely aware of the
entire universe He creates and fully appreciates the mathematical beauty of its
laws of physics that He uses in His creation, this omniscient conscious
awareness could have enormous value and help make the entire world, including
God and His own sentient experiences, the best possible world.  Cancer and
earthquakes may be logical consequences of these laws of physics.  God Himself
may grieve over the evils that are a consequence of the laws of physics that
give Him even much greater joy, but there may be this inevitable trade-off.

\newpage

Note that I am not saying that God is constrained by the laws of physics in the sense that they are any external limitation on His happiness.  But if He had chosen to eliminate the evil, the laws of physics that He would have had to use to do that would have been different from what they actually are.  Quite possibly they would have had to be less elegant and beautiful to Him than the actual laws He did choose, and such less-orderly laws may well have made Him much less happy, so that the total value of the world could well have been reduced.
	
One might complain that the choices God make appear selfish, placing His happiness above that of His creatures.  But if God's choice really did maximize the total happiness of the world, then I see it as quite justified.  Who are we to complain when God does what He pleases, especially if what He does actually maximizes the total intrinsic value of the world?
	
What I think is wrong about most human selfishness, including much of what I see in myself, is that it places a higher value on one's own happiness than on that of others.  Far too often we act to increase our happiness by a certain amount, whereas if we had acted differently, other people's happiness could have increased many-fold.  For example, if those of us who are well-off in developed nations contributed even just 10\% of our income to help those less well-off in poorer nations, our personal happiness might go down by a tiny fraction, but I strongly suspect it would go up by a much larger fraction for those helped.  I am not saying that it is wrong to enjoy happy experiences and to seek to increase them, so long as doing that does not cause a greater decrease in happiness in others, but I am saying that very often it would be better to sacrifice a bit of one's personal happiness to give a greater increase in the happiness of others.
	
Therefore, if it is indeed true, as I postulate, that God gets tremendous satisfaction from the elegant laws of physics that He creates, it would not be right to expect Him to give that up for a much smaller happiness that might be afforded His creatures if they were spared from evils like cancer and earthquakes.  God has created this universe to lead to much human joy, and I suspect that He has indeed sacrificed much personal happiness to achieve that.  In particular, I believe that Jesus greatly reduced the happiness He otherwise would have had in Heaven by coming to earth as a human and enduring the suffering and shame He did, especially when He was unjustly executed on the Cross.  But I think it would be unfair for us to expect God to give up more happiness than we would gain by some action of His, such as perhaps changing His usual laws of physics to prevent cancer and earthquakes.
	
Even if one believed in human libertarian free will in the incompatibilist sense as the cause for other evils, much of my discussion would still apply to so-called natural evils (the negative intrinsic value of the sentient experiences of suffering attributed to cancer, earthquakes, etc., and not to human choices, though even with these examples one must exclude human choices to smoke or to build inadequate buildings in earthquake-prone regions).  So I think that readers who do not see problems reconciling libertarian free will with creation from nothing might still wish to consider my speculations as to why God seems to follow the same laws of physics most of the time.
	
However, for those of us who simply do not find libertarian creaturely free will plausible (since to me it seems logically inconsistent to assume that God can create from nothing what He does not completely determine, and since I do believe that God has created from nothing absolutely everything other than Himself and other than entities that I believe can neither be created nor destroyed, such as logically necessary truths like mathematical theorems), I would suggest that these speculations might help explain human evil as well.  If our actions are determined by God, this may also inevitably lead to evils produced by humans (but with the ultimate determination going back to God).
	
I am not saying that we are forced to do evil purely because of our genes, since genes are only a small part of the causal chain.  There is far more information in what might be called nurture rather than nature, though our genes help form recurring traits, such as personality.  For example, the tone of my writing may partially reflect genetic influences on my personality traits that persist, but the content of what I write is much more influenced by what I have learned.
	
In the Everett `many worlds' version of quantum theory, a person is continually branching into many copies (each copy in a different Everett `world,' which should not be confused with the entire world of all that exists).  Even with exactly the same genes and previous experiences (the same `nature' and `nurture'), the outcomes in the different Everett `worlds' will be different.  In ours Hitler was an evil monster.  But I suspect that in most Everett `worlds' with the same early `nature' and `nurture' for Hitler, he was not nearly so evil.  (Of course, there is the flip side:  in most Everett `worlds,' Mother Teresa presumably also did not turn out so good as she did in ours.)
	
I believe that it is a consequence of the laws of physics that when a person is faced with a moral choice, in some Everett `worlds' in which that choice is made, an evil choice is made, one that reduces the total happiness of the conscious beings in that Everett `world.'  There will also be Everett `worlds' in which a good choice is made, which increases total happiness.  (One might postulate that Jesus was an exception, choosing to incarnate Himself with no quantum amplitude to make any evil choices.)  
	
If God had chosen to use sufficiently different laws of physics (perhaps even just to the extent of collapsing the quantum state appropriately as needed), I think He could have eliminated not only all natural evil, but also all human evil.  However, that might actually have decreased the total happiness of the entire world by reducing His own happiness greatly at using less elegant laws of physics.

This consideration leads into the theological argument for an Everett multiverse.  It seems highly plausible that leaving the quantum state uncollapsed is mathematically much more elegant and beautiful than collapsing it as would seem to be needed to avoid superpositions of macroscopically different situations (different `Everett worlds').  This in fact is the main reason I personally favor the uncollapsed Everett hypothesis about the quantum state.  In my {\it Optimal Argument for the Existence of God}, I am assuming that a suitable omniscient God would greatly appreciate the elegance of the universe and prefer a more elegant version (such as Everett's) over a less elegant one, such as one with `random' collapses of the quantum state.

But if it did not reduce God's happiness very much for Him to collapse the quantum state, it would seem that He would do so in ways to increase creaturely happiness under this new assumption that it would not cost God more than the gain to us (an alternative to what I actually believe), so then it would seem that God would indeed collapse the quantum state in such ways to increase creaturely happiness, which apparently could be enormously increased by suitable collapses of the quantum state.  Of course, one cannot rule out some other reason why God might choose to collapse the quantum state in ways that do not greatly increase creaturely happiness over that in our actual world, but it would certainly seem mysterious to me why an omnibenevolent God would not choose collapses to increase creaturely happiness greatly if that would be at less cost to His own happiness.  The fact that He has apparently not done so suggests that there is a good reason why He has not, such as for maintaining the high level of pleasure He gets from very elegant laws of physics that do not have collapses of the quantum state.

Another alternative is that conceivably there is an elegant way for the quantum state to be collapsed, and God uses and greatly appreciates that elegant way, even though it does not increase creaturely happiness nearly so much as what seems logically possible from a suitably different set of collapses of the quantum state to avoid creaturely suffering and maximize creaturely happiness.  For example, suppose somehow there were some elegant set of discrete times at which the quantum state were to be collapsed, as well as an elegant discrete set of basis states at each of those times into which the quantum state could collapse.  (It is not obvious how to make an elegant choice for either of these, but for the sake of argument let us suppose God has made some particular choices for them.)  Then one might make the {\it ad hoc} postulate that at each of the collapse situations, God takes the two basis states that give the largest absolute value of the inner product with the actual quantum state before the collapse (with the basis state giving the largest value labeled 0 and the basis state giving the next largest value 1; it would be a set of zero cases, which we can ignore, in which there is a tie for the largest two values), and then God collapses the quantum state onto whichever one of those two basis states corresponded to the next binary digit of $\pi$.

This at least is a definite algorithm for which of the discrete basis states to collapse the quantum state onto at each of the discrete times, though it leaves aside the question of how to choose the basis states and the times.  The algorithm is fairly simple, being described by the single final sentence of the previous paragraph, so in some sense it might be regarded as fairly elegant, though to me is certainly does not look so elegant as simply leaving the quantum state uncollapsed.  Furthermore, once one completes the algorithm by giving a specification of the discrete times for the collapses and the of the basis states at those times, the algorithm would presumably be even more complicated and less elegant.  Therefore, even though it seems logically possible that God could collapse the quantum state in an elegant way (which most likely would not much increase the ratio of creaturely happiness to sadness over that in the Everett version of quantum theory with no collapses, which seems consistent with our observations), it does seem to run the danger of making the universe less elegant and less pleasurable for God.

Thus it seems simpler to me to focus on the two extreme cases:  (1)  God collapses the quantum state to maximize creaturely happiness.  (2)  God does not collapse the quantum state at all.  Since (1) certainly seems contrary to our observations, (2) appears to me much more plausible.  It might maximize total happiness once God's appreciation of the elegance of the universe is included.

Now of course I cannot rule out intermediate cases, such as the one described above in which God does use a fairly elegant way to collapse the quantum state.  Another alternative that I find much more plausible is that most of the time God does not collapse the quantum state, but on certain very special occasions, such as the Resurrection of Jesus Christ, and perhaps in certain other miracles, He does choose to do so for reasons that override His usual desire for elegance.  However, it seems to me that most of the time, God does not choose to collapse the quantum state in a way that obviously appears to have much moral significance for us creatures within the universe, so when applying theological as well as philosophical and scientific considerations to our observations, it seems most consistent with my assumptions and with our observations that God rarely if ever collapses the quantum state.

In summary, I see not only philosophical and scientific but also theological reasons for an Everett multiverse, for the reality of different quantum outcomes to choices that the universe faces.

I might also end by including the final four paragraphs that I wrote in an earlier essay on {\em The Superb Universe} in {\em Faith Seeking Understanding: Essays in Memory of Paul Brand and Ralph D.\ Winter}, edited by David Marshall (William Carey Library, Pasadena, California, 2012):

Of course, there might also be other reasons as well for why God values the way He actually runs the world.  Often justice is viewed as a reason God acts and sometimes reduces individual human happiness through punishment.  However, I tend to see justice as a way to maximize total happiness, such as preventing thieves from gaining a small degree of happiness by stealing from victims whose happiness is typically reduced by a much greater degree by their loss of goods.  So it is hard for me to see justice as an independent reason for God's choices, but it is conceivable that God Himself gets satisfaction from some sense of justice that does not merely have the goal of maximizing happiness within the universe He has created.
	
Even if my speculative hypothesis can give an intellectual explanation for how the existence of an all-good, all-knowing, and all-powerful God can be consistent with the existence of evil, I do not wish to imply that we should passively sit back and accept the continuation of the evil we observe.  I believe that God has created this universe so that we can join with Him in working to reduce evil and suffering.  I do not think we can take pride in what I regard as the myth that we of our own independent free will choose to do good and fight against evil.  But I do believe that by God's grace, part of the way He determines our actions is by His call to join Him in the creation of happiness and joy.  Even when following this call may lead to a reduction of one's own happiness, a true personal sacrifice, we should follow that call to give an even greater increase in the happiness of others.  I believe that it is the existence of this call by God, and the extreme example of carrying it out by His Son Jesus, that helps make this world the best possible.
	
In conclusion, I do not claim to have found the only, or even the main, reason why God has created a universe in which there is evil and suffering.  But as a physicist with a dim sense of the enormous elegance and beauty of the laws of physics, it seems plausible to me that God may value this elegance (which we only partially grasp by virtue of having been created in the image of God) to such an extent that His faithfully employing the same orderly laws over and over really does maximize the goodness of the total world, despite the local evils and sufferings that result.

My speculation is that this is the best possible world, and that one aspect that helps make it the best is God's appreciation for elegant laws of physics.  This speculation is of course highly tentative, uncertain, and controversial.  There are many open issues in this conjectural theodicy, such as what the value of a world is, why elegant laws of physics might have value to God, whether there even is a best possible world, etc.  But as a tentative working hypothesis, I believe God has created the universe as a Superb Design, and that His creation is literally for the best.

\end{document}